\documentclass[cernpreprint,USenglish]{na61doc}
\usepackage[utf8]{inputenc}
\usepackage[T1]{fontenc}
\usepackage{amsmath}
\usepackage{url}
\usepackage{multirow}
\usepackage{colortbl}
\usepackage[colorlinks=true,linkcolor=firebrick,citecolor=darkgreen,urlcolor=darkblue]{hyperref}
\usepackage{mathptmx} 
\usepackage{enumerate}
\usepackage{lineno}
\usepackage{xspace}
\usepackage{color}
\usepackage{tikz}
\usetikzlibrary{patterns}
\usepackage{varwidth}
\usepackage{placeins}
\usepackage{bbding}
\usepackage{enumerate}
\usepackage{afterpage}
\usepackage{longtable}
\usepackage{varwidth}
\usepackage{flafter}
\usepackage{cite}

\usepackage{comment}

\renewcommand{\NASixtyOne}{\mbox{NA61/SHINE}\xspace}

\newcommand{\pp}{\mbox{\textit{p}+\textit{p}}\xspace}

\newcommand{\GeantFour}{{\scshape Geant4}\xspace}

\newcommand{\EposLong}{{\scshape Epos1.99}\xspace}
\newcommand{\FTFPBERT}{{\scshape Ftfp\_Bert}\xspace}

\newcommand{\MeV}{\mbox{Me\kern-0.1em V}\xspace}
\newcommand{\GeV}{\mbox{Ge\kern-0.1em V}\xspace}
\newcommand{\TeV}{\mbox{Te\kern-0.1em V}\xspace}
\newcommand{\A}{\textit{A}\xspace}
\newcommand{\GeVc}{\mbox{Ge\kern-0.1em V\kern-0.15em /\kern-0.05em\textit{c}}\xspace}
\newcommand{\MeVc}{\mbox{Me\kern-0.1em V\kern-0.15em /\kern-0.05em\textit{c}}\xspace}

\newcommand{\AGeVc}{\A~\GeVc}
\newcommand{\Ar}{\textsuperscript{40}Ar\xspace}
\newcommand{\Sc}{\textsuperscript{45}Sc\xspace}
\newcommand{\Xe}{\textsuperscript{129}Xe\xspace}
\newcommand{\La}{\textsuperscript{139}La\xspace}
\newcommand{\Pb}{\textsuperscript{208}Pb\xspace}
\newcommand{\ArSc}{\mbox{\Ar~+~\Sc}\xspace}
\newcommand{\XeLa}{\mbox{\Xe~+~\La}\xspace}
\newcommand{\PbPb}{\mbox{\Pb~+~\Pb}\xspace}
\newcommand{\dedx}{\mbox{\ensuremath{\textrm{d}E\!/\!\textrm{d}x}}\xspace}
\newcommand{\sNN}{\ensuremath{\mkern-4mu\sqrt{s_{\mathrm{NN}}}}\xspace}

\newcommand{\pt}{\ensuremath{p_{\textrm T}}\xspace}

\newcommand{\mt}{\ensuremath{m_{\textrm T}}\xspace}
\newcommand{\y}{\ensuremath{{y}}\xspace}
\newcommand{\dd}{\ensuremath{{\textrm d}}\xspace}

\definecolor{darkred}{rgb}{0.5,0,0}
\definecolor{darkblue}{rgb}{0,0,0.5}
\definecolor{firebrick}{rgb}{0.75,0.125,0.125}
\definecolor{darkgreen}{rgb}{0,0.5,0}
\definecolor{redShading}{RGB}{229,127,127}
\definecolor{colorPSDCentral}{RGB}{66,129,164}
\definecolor{colorPSD150}{RGB}{72,169,166}
\definecolor{colorPSD75}{RGB}{212,180,131}
\definecolor{colorPSD19}{RGB}{193,102,107}

\ShineTitle{
 $\pmb{K^*(892)^0}$ production and the time between freeze-outs in \ArSc collisions by \NASixtyOne at the CERN SPS
 }

\PreprintIdNumber{CERN-EP-2026-187}

\ShineJournal{Eur. Phys. J. C (Letter)}

\ShineJournalRef{}
\ShineDOI{}

\ShineAbstract{
The analysis of the production of strange $K^{*}(892)^0$ resonances allows us to better understand the temporal evolution of high-energy nucleus--nucleus collisions. In particular, the ratio of $K^{*}(892)^0$ to charged kaon yields is used to determine the time interval between chemical and kinetic freeze-outs. In this paper, the first measurements of $K^{*}(892)^0$ production in central \ArSc collisions at the CERN Super Proton Synchrotron are reported. They were performed by \NASixtyOne at collision center-of-mass energies per nucleon pair \sNN $=$ 8.8, 11.9, 16.8 \GeV. 
The obtained $\langle K^{*}(892)^0 \rangle/\langle K^{+} \rangle $ and $\langle K^{*}(892)^0 \rangle/\langle K^{-} \rangle$ mean multiplicity ratios are compared with corresponding results in \pp collisions, allowing for an estimate of the time interval between chemical and thermal freeze-outs in the \ArSc system. These are the first such results reported for \ArSc collisions.
}

\begin{document}

\maketitle

\section{Introduction}

Resonance production is one of the fundamental observables to study the intricate dynamics of high-energy collisions. In high-density systems created in nucleus--nucleus collisions, the properties (for example, widths, masses, branching ratios) of some of them are expected to be modified due to the partial restoration of chiral symmetry~\cite{Pisarski:1981mq, Brown:1991kk, Brown:1995qt, Milov:2008dd}. The resonance spectra and yields are also indispensable inputs for hydrodynamic-inspired Blast-Wave parameterizations and Hadron Resonance Gas frameworks~\cite{Schnedermann:1993ws, Becattini:2005xt}. Finally, the analysis of the strange $K^{*}(892)^0$ resonances allows us to better understand the temporal evolution of high-energy nucleus--nucleus collisions. More specifically, the yield ratio of $K^{*}(892)^0$ to charged kaons is used to determine the time interval between chemical freeze-out (the end of inelastic interactions, or more precisely, the fixing of quark composition) and thermal freeze-out, also called the kinetic one (the end of elastic interactions)~\cite{Markert:2002rw}. The lifetime of the $K^{*}(892)^0$ resonance ($\approx 4$ fm/$c$~\cite{ParticleDataGroup:2024}) is comparable to the expected duration of the rescattering hadronic gas phase between the two freeze-out stages. Consequently, a certain fraction of $K^{*}(892)^0$ resonances will decay inside the fireball medium. The momenta of their decay products are expected to be substantially modified by elastic scatterings, precluding the experimental reconstruction of the resonance via an invariant mass analysis. Under these circumstances, a suppression of the observed $K^{*}(892)^0$ yield is expected. The reduction of $K^{*}/K$ yield ratio ($K^{*}$ stands for $K^{*}(892)^0$ and/or 
$\overline{K^{*}}(892)^0$ and $K$ denotes $K^{+}$ and/or $K^{-}$) with increasing system size was indeed observed at SPS (Super Proton Synchrotron), RHIC (Relativistic Heavy Ion Collider), and LHC (Large Hadron Collider) energies, and for central Pb+Pb or Au+Au collisions, the estimated time intervals between freeze-outs are at the level of a few fm/$c$ \cite{NA61SHINE:2020czr, STAR:2022sir, Sahoo:2022qkl, ALICE:2023edr, Das:2022lqh}.

This paper presents the first measurements of $K^{*}(892)^0$ spectra and yields obtained for collisions of intermediate-mass nuclei, \ArSc, at three SPS energies (\sNN $=$ 8.8, 11.9, 16.8 \GeV). The $K^{*}(892)^0$ multiplicities, compiled with published results on $K^{*}(892)^0$, $K^{+}$, and $K^{-}$ production in \pp collisions at similar energies~\cite{NA61SHINE:2020czr, NA61SHINE:2021wba, NA61SHINE:2017fne}, as well as $K^{+}$ and $K^{-}$ yields in \ArSc collisions at the same energies~\cite{NA61SHINE:2023epu} allowed to estimate the time intervals between freeze-outs in \ArSc collisions.
\section{The \NASixtyOne detector}
\label{sec:detector}

Data used for the measurements were recorded by the \NASixtyOne \cite{NA61:2014lfx} multipurpose, fixed-target experiment located at the CERN SPS. The main components of the detection system used in this analysis are four large-volume Time Projection Chambers (TPCs). Two of them (Vertex TPCs, denoted as VTPC-1/2) are located downstream of the target inside superconducting dipole magnets. Two others (Main TPCs, denoted as MTPC-L/R) are located downstream of the magnets. The Projectile Spectator Detector (PSD), a hadronic calorimeter positioned downstream of the MTPCs, is used for centrality determination. Primary beams of fully ionized \Ar nuclei were extracted from the SPS accelerator. The three beam position detectors (BPDs), placed upstream of the target, precisely measured individual beam particle trajectories. The target consisted of \Sc plates installed upstream of VTPC-1. The full description of detector components used for \ArSc data taking can be found in Refs.~\cite{NA61SHINE:2021nye, NA61SHINE:2023gez, NA61SHINE:2023epu, SHINE:2024xtq, NA61SHINE:2025whi}.
\section{Analysis}
\label{sec:analysis}

The aim of this analysis was to obtain results on inclusive $K^*(892)^0$ production in 0--10\% central \ArSc collisions at \sNN $=$ 8.8, 11.9, 16.8 \GeV. In \NASixtyOne, central collisions are selected by requiring a low value of the forward energy -- the energy emitted into the region occupied by the projectile spectators~\cite{NA61SHINE:2021nye}. The analysis procedures used in this work were largely inherited from the two previous \NASixtyOne $K^*(892)^0$ analyses of \pp interactions described in Refs.~\cite{NA61SHINE:2020czr, NA61SHINE:2021wba}. The details specific for $K^*(892)^0$ analysis in \ArSc collisions can be found in Ref.~\cite{BKozlowski_PhD}.

The results presented in this paper are based on \ArSc events recorded in 2015 at three beam momenta: 40\A, 75\A, and 150\AGeVc, corresponding to \sNN $=$ 8.8, 11.9, and 16.8 \GeV, respectively. A set of event cuts was applied to select good-quality \ArSc interactions. (Semi)central collisions were selected on-line by the trigger logic (see Refs.~\cite{NA61SHINE:2021nye, NA61SHINE:2023gez, NA61SHINE:2023epu, SHINE:2024xtq, NA61SHINE:2025whi} for details). Additionally, off-line cuts were implemented to select the 10\% of \ArSc collisions with the lowest energy values measured by a subset of PSD modules (see Refs.~\cite{NA61SHINE:2021nye, NA61SHINE:2023epu, SHINE:2024xtq} for methodology details). The kinematic acceptance of the PSD can be found in Ref.~\cite{ref_PSD_acc}. The beam particle (fully ionized \Ar nucleus) was measured using BPD-3 in combination with either BPD-2 or BPD-1. No off-time beam particles were allowed within a time window of $\pm 4$ $\mu$s around the trigger (beam) particle. No interaction-event triggers were allowed within a time window of $\pm 25$ $\mu$s around the trigger particle. The $z$ position (along the beam line) of the properly fitted primary \ArSc interaction vertex was required to be no further than 2 cm away from the center of the \Sc target. After applying the above selection criteria, the numbers of accepted events were 1.29 $\times$ 10$^6$, 1.16 $\times$ 10$^6$, and 3.94 $\times$ 10$^5$ for 40\A, 75\A, and 150\AGeVc, respectively.

The $K^*(892)^0$ resonance production was analyzed via its $K^{+}\pi^{-}$ dominant decay mode. A set of track quality cuts was applied to individual daughter tracks. The total number of reconstructed TPC clusters on the track was required to be at least 30. The sum of reconstructed clusters in VTPC-1 and VTPC-2 was at least 15. The distance between the track extrapolated to the interaction plane and the interaction point (the so-called impact parameter) was at most 4~cm in the horizontal (bending) plane and 2~cm in the vertical (drift) plane. The track's total momentum (in the laboratory reference frame) was required to be at least 3~\GeVc. The track's transverse momentum\footnote{Transverse momentum (\pt) is the component of the total momentum perpendicular to the beam axis.} was required to be at most 1.5~\GeVc. The candidates for kaons and pions were selected based on their energy loss (\dedx) in the gas volume of TPCs. The ranges $\pm 2$ $\sigma_{\pi}$ (for pions) and $\pm 1.5$ $\sigma_{K}$ (for kaons) around their empirical parametrizations of Bethe-Bloch curves were selected. The $\sigma$ quantity represents a typical standard deviation of a Gaussian function fitted to the \dedx distribution of charged pions ($\sigma_{\pi}$ $=$ 0.052) or kaons ($\sigma_{K}$ $=$ 0.044)~\cite{NA61SHINE:2020czr, NA61SHINE:2021wba}. Finally, the opening angle between momentum vectors of the $K^{+}$ and $\pi^{-}$ candidates (in laboratory reference frame) was at least 2$^\circ$ for 40\AGeVc, 1.5$^\circ$ for 75\AGeVc, and 1$^\circ$ for 150\AGeVc.

\begin{figure}[h]
\centering
\includegraphics[width=0.5\textwidth]{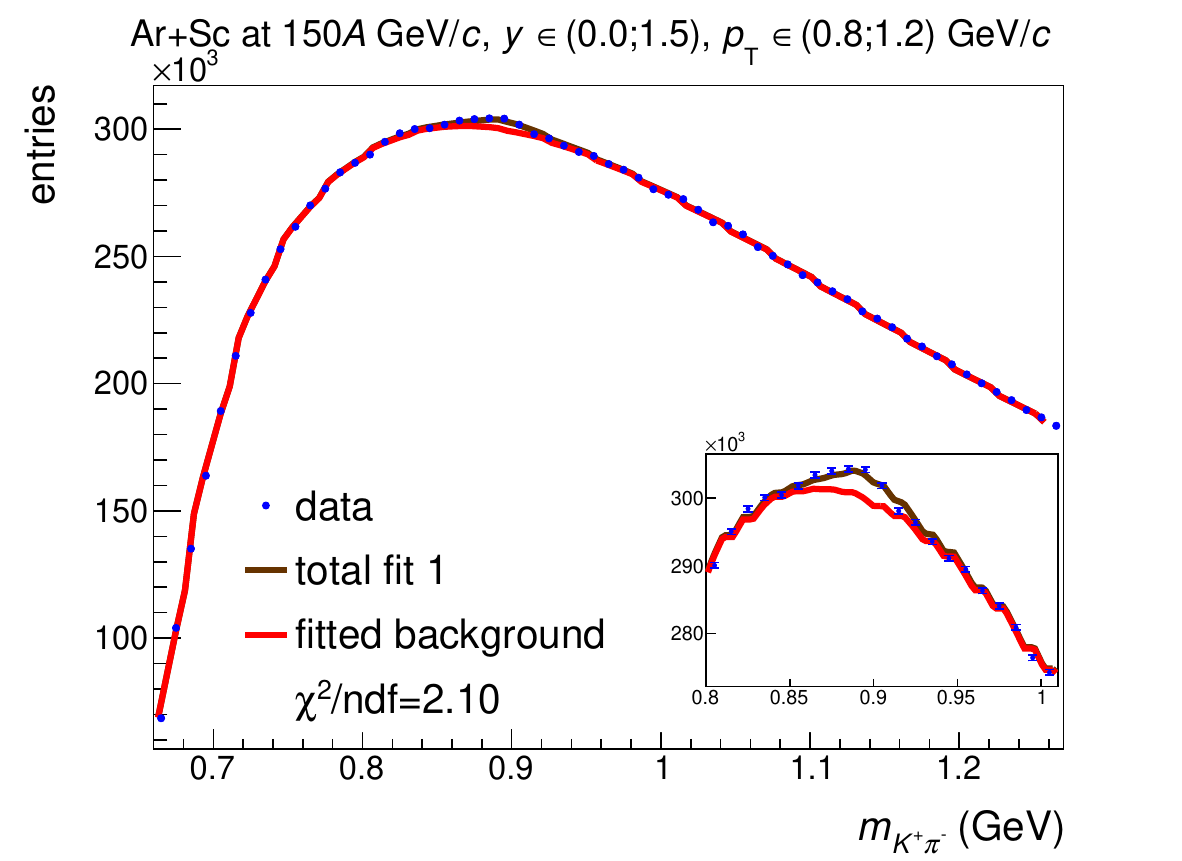}
\includegraphics[width=0.49\textwidth]{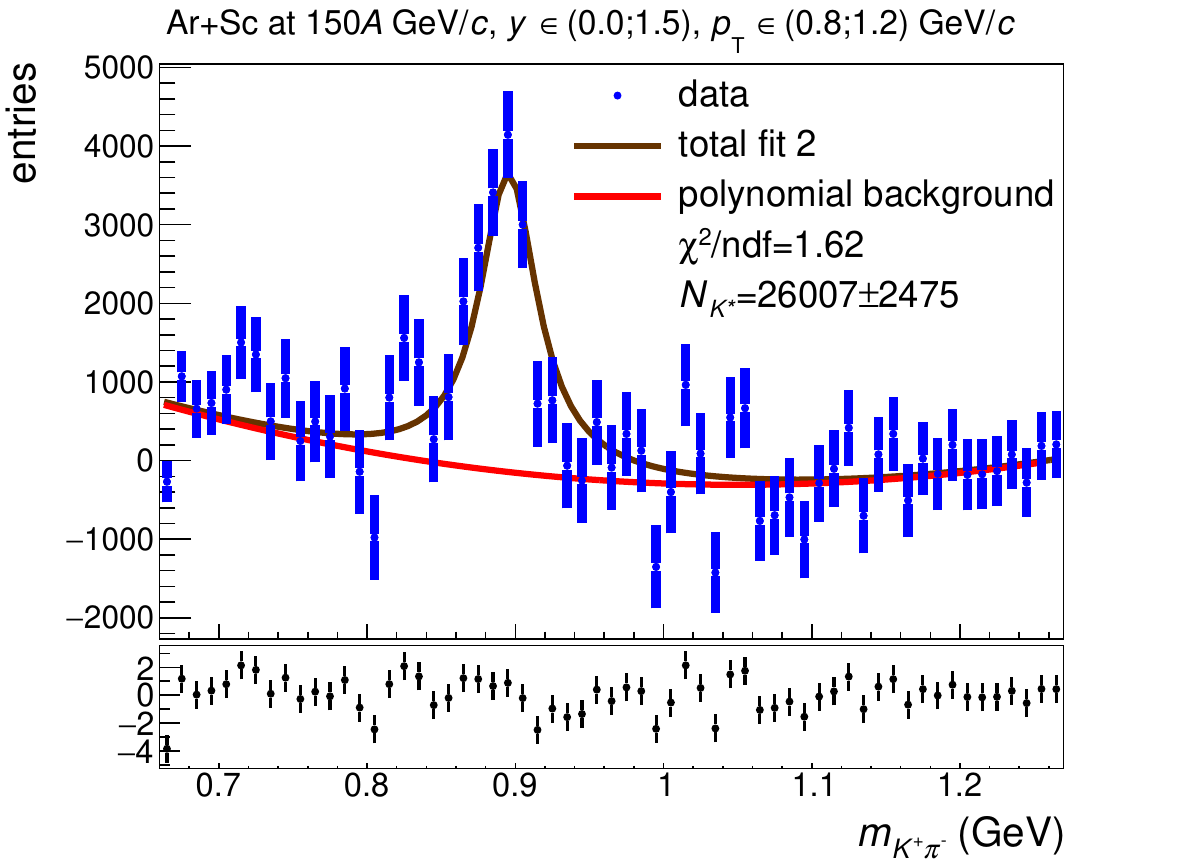}
\caption{\textit{Left}: An example of invariant mass spectrum (blue dots) fitted with Eq.~\eqref{eq:fit} (brown line, visible near the $K^{*}(892)^{0}$ resonance mass). The red line shows the \textit{fitted background} estimated using the template method (reconstructed Monte Carlo from \EposLong~\cite{Werner:2005jf}, as well as mixed events). 
\textit{Right}: An invariant mass spectrum after \textit{fitted background} subtraction. The red line is the residual background described by a second-order polynomial curve. The bottom panel presents the normalized residuals: the differences between the brown curve (\textit{total fit 2}) and the blue points, divided by their uncertainties.}
\label{fig:invmass}
\end{figure}

After the selection of events and tracks, the invariant mass distributions of $K^{+}\pi^{-}$ candidate pairs were obtained. In this analysis the $K^*(892)^0$ signal was extracted using the \textit{template method}~\cite{NA61SHINE:2020czr, NA61SHINE:2021wba}, which for $K^*(892)^0$ meson was found to allow a more precise background subtraction than the \textit{standard} procedure, based on mixed events\footnote{In the \textit{standard (mixing) method} the combinatorial background is estimated by invariant mass spectra calculated for $K^{+}\pi^{-}$ pairs originating from different events. } only~\cite{NA61SHINE:2020czr}. In the template method, the invariant mass spectra of $K^{+}\pi^{-}$ candidate pairs are fitted with the function:
\begin{equation}
f(m_{K^+\pi^-}) = a \cdot T_{\mathrm{res}}^{\mathrm{MC}}(m_{K^+\pi^-}) + b \cdot T_{\mathrm{mix}}^{\mathrm{DATA}}(m_{K^+\pi^-})+ c \cdot BW(m_{K^+\pi^-})\, ,
\label{eq:fit}
\end{equation}
where $T_{\mathrm{res}}^{\mathrm{MC}}(m_{K^+\pi^-})$ is the resonance background estimated using reconstructed Monte Carlo events (combination of tracks that come from decays of resonances different than $K^{*}(892)^{0}$ and combination of tracks where one comes from the decay of a resonance and one comes from direct production in the primary interaction), $T_{\mathrm{mix}}^{\mathrm{DATA}}(m_{K^+\pi^-})$ is background estimated using the event-mixing method ($K^{+}$ and $\pi^{-}$ candidates originating from different events), $BW(m_{K^+\pi^-})$ is the non-relativistic Breit-Wigner distribution, and $a$, $b$, and $c$ are normalization factors (see Ref.~\cite{NA61SHINE:2021wba} for details). The examples of invariant mass spectra are presented in Fig.~\ref{fig:invmass}. In the fitting procedure, fixed values~\cite{ParticleDataGroup:2024} of $K^{*}(892)^{0}$ mass ($m_0 = $ 895.55~\MeV) and width ($\Gamma = $ 47.3~\MeV) were used in the $BW$ shape. The \textit{total fit 1}, shown in Fig.~\ref{fig:invmass} (\textit{left}), represents the fit given by Eq.~\eqref{eq:fit}, whereas the subsequent \textit{total fit 2}, presented in Fig.~\ref{fig:invmass} (\textit{right}), denotes a combination of the \textit{BW} and a second-order polynomial curve, fitted to the data after subtraction of the \textit{fitted background}.

For each selected transverse momentum and rapidity\footnote{Rapidity (\y) is defined as $\y = \frac{1}{2} \ln \frac{E+p_z}{E-p_z}$, where $E$ is the total energy of the particle and $p_z$ is the component of total momentum longitudinal to the beam axis. In this paper, all rapidity values are given in the collision center-of-mass frame, where the \textit{mid-rapidity} region corresponds to particles with $\y \approx 0$.} bin the \textit{raw} number of $K^*(892)^0$ particles ($N_{K^*}$) was computed as the integral (divided by the bin width) over the $BW$ signal of \textit{total fit 2} in Fig.~\ref{fig:invmass} (\textit{right}) (see Ref.~\cite{NA61SHINE:2021wba} for details). The raw $K^*(892)^0$ numbers were then corrected for geometrical acceptance, detector and reconstruction inefficiencies as well as biases due to event and track selection procedures using \EposLong events processed through the full detector simulation (\GeantFour~\cite{GEANT4:2002zbu}) and the standard reconstruction chain. Corrections were determined by comparing the normalized numbers of generated $K^*(892)^0$ mesons with those selected from the reconstructed Monte Carlo. The corrections were calculated separately for \pt spectra integrated over rapidity and rapidity spectra integrated over \pt. The binning is given in Sec.~\ref{sec:results}. The final yields are also corrected for $K^{+}$ and $\pi^{-}$ identification inefficiency and branching ratio of $K^*(892)^0$ decay into $K^{+}\pi^{-}$~\cite{NA61SHINE:2021wba}.
Systematic uncertainties were calculated~\cite{BKozlowski_PhD} by varying event and track selection criteria, modifying the signal extraction procedure, as well as changing the model used to generate Monte Carlo events (from \EposLong to \FTFPBERT~\cite{Allison:2016lfl}).

\section{Results}
\label{sec:results}

\begin{figure}[h]
    \centering
    \begin{minipage}{0.45\textwidth}
        \centering
        \includegraphics[width=\textwidth]{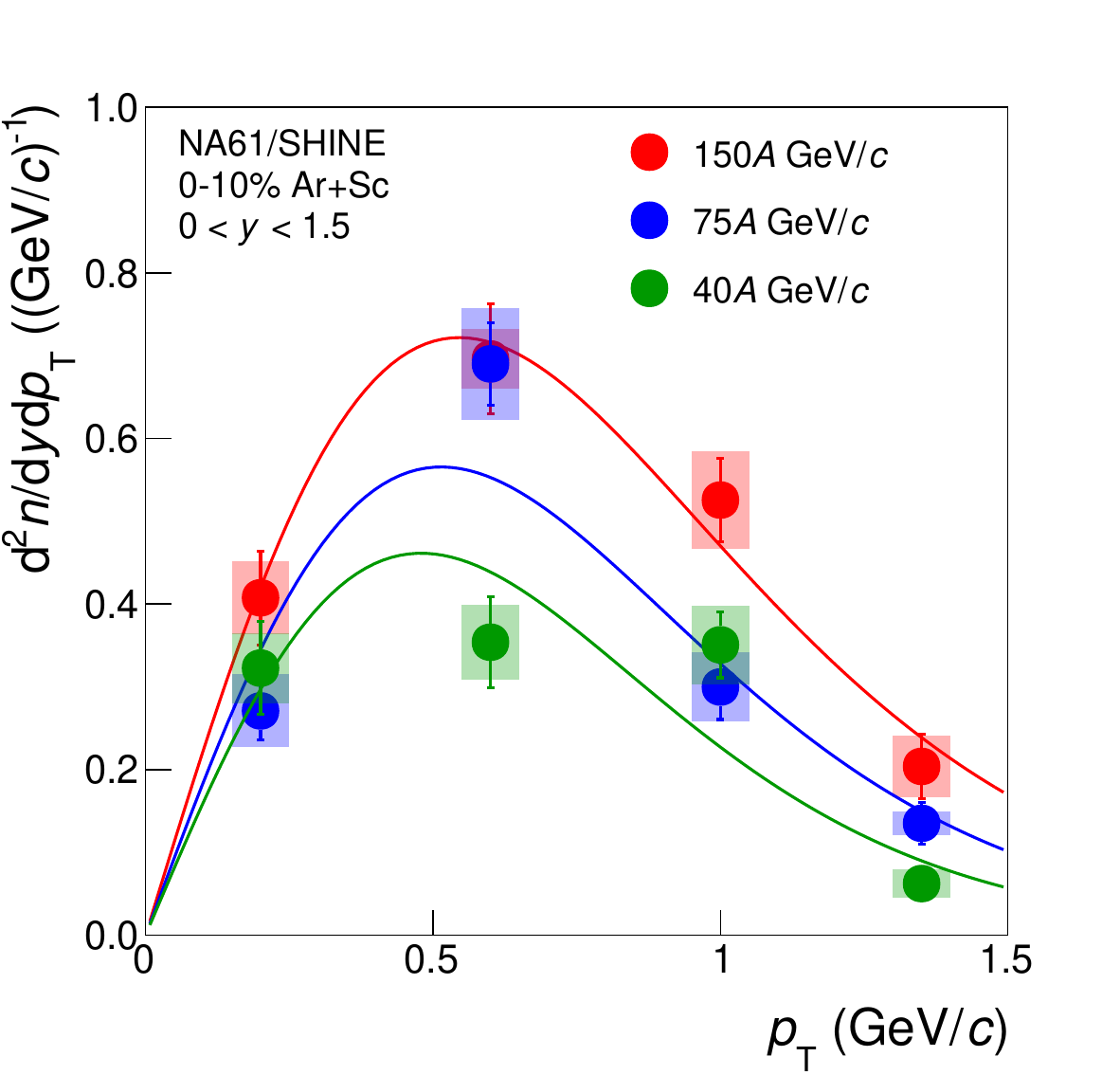}
    \end{minipage}
    \begin{minipage}{0.45\textwidth}
        \centering
        \includegraphics[width=\textwidth]{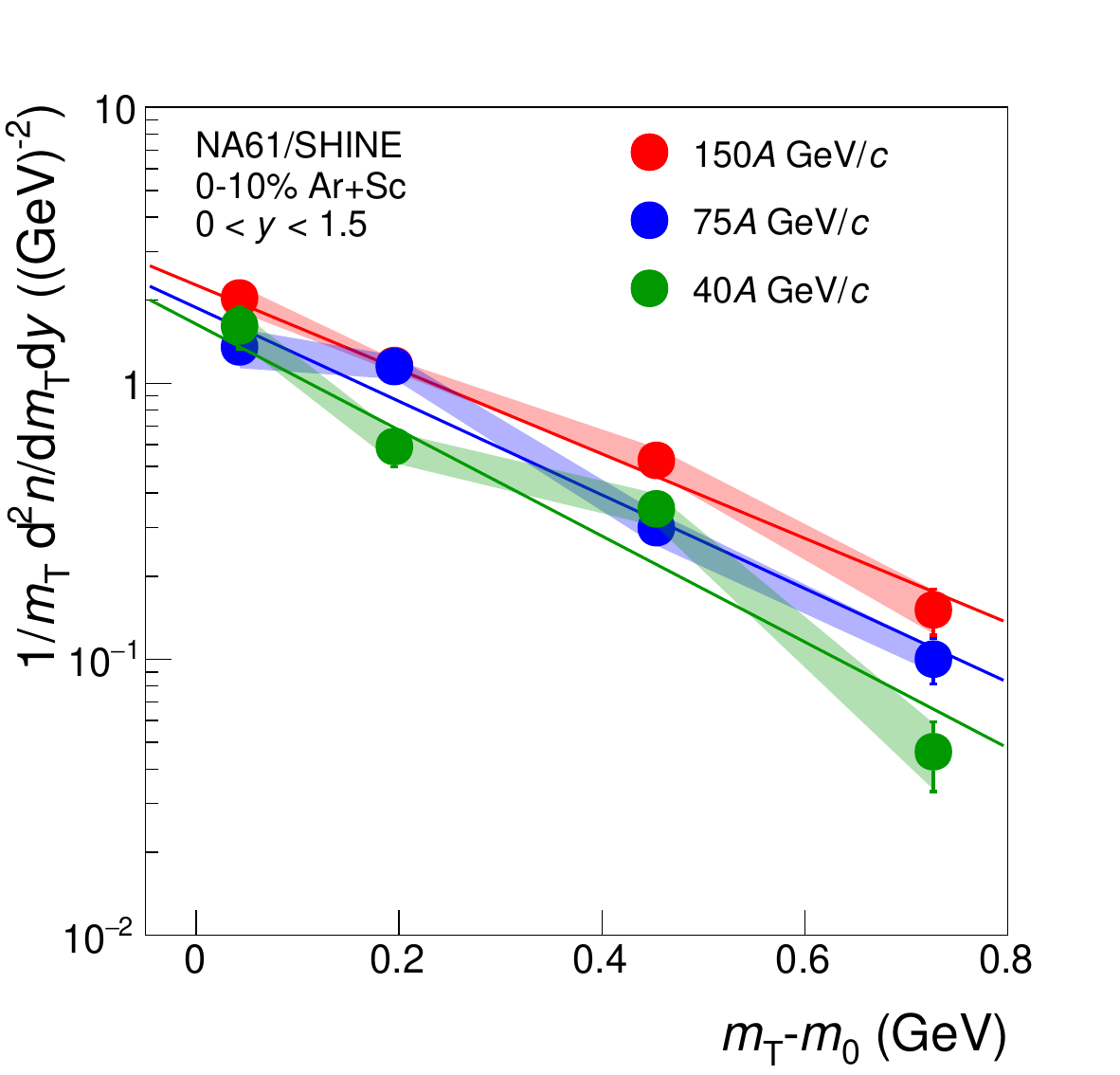}
    \end{minipage}
    \begin{minipage}{0.45\textwidth}
        \centering
        \includegraphics[width=\textwidth]{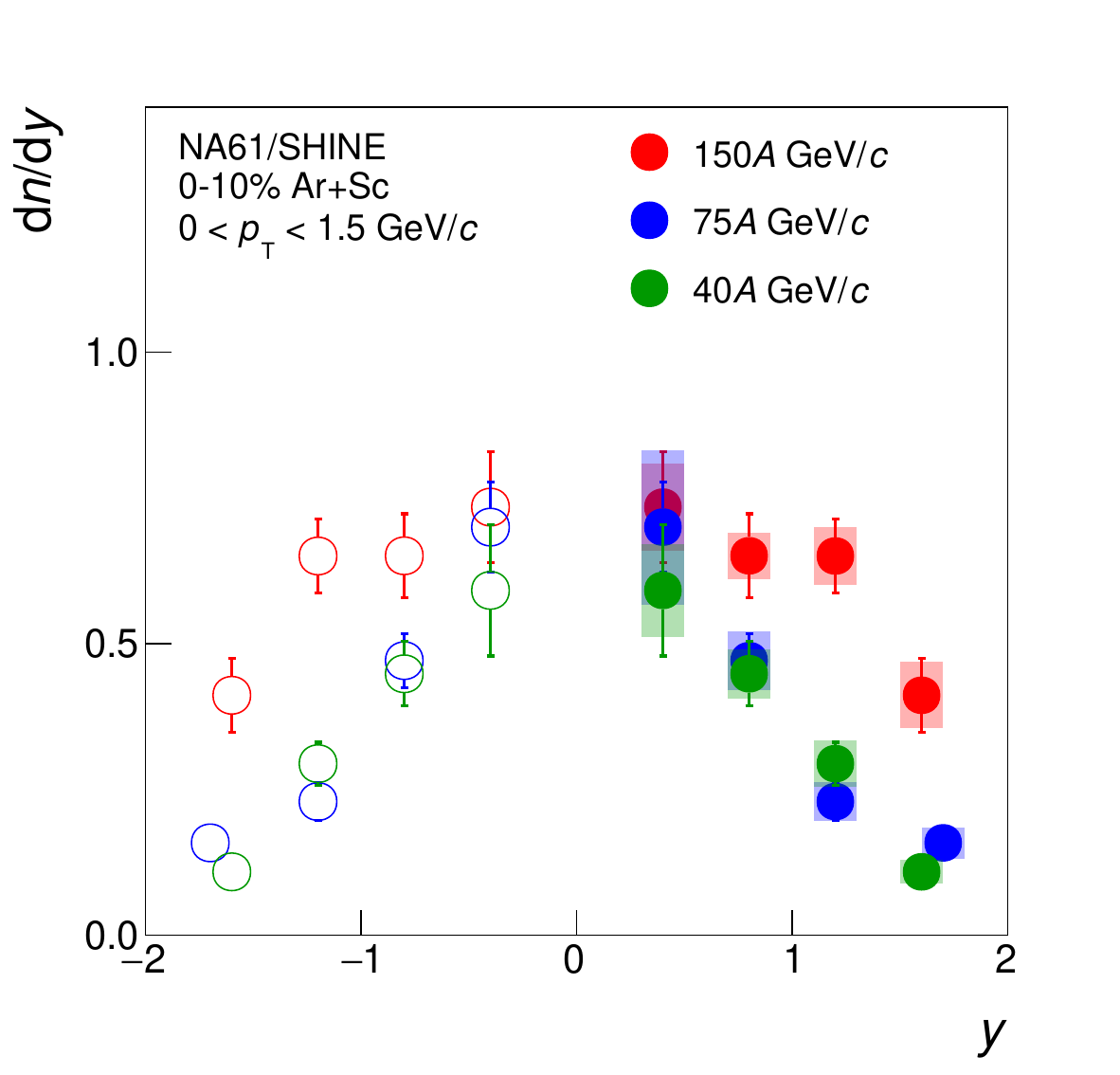}
    \end{minipage}
    \begin{minipage}{0.45\textwidth}
        \centering
        \caption{The transverse momentum (\textit{top left}), transverse mass (\textit{top right}), and rapidity (\textit{bottom left}) spectra of $K^*(892)^0$ mesons produced in 0--10\% central \ArSc collisions at 40\A, 75\A, and 150\AGeVc (\sNN $=$ 8.8, 11.9, 16.8 \GeV). Bars denote statistical uncertainties, color bands or boxes -- systematic ones.}
        \label{fig:ptmty}
    \end{minipage}
\end{figure}

In this section the \NASixtyOne results on $K^{*}(892)^0$ meson production in 0--10\% central \ArSc collisions at 40\A, 75\A, and 150\AGeVc (\sNN $=$ 8.8, 11.9, 16.8 \GeV) are presented. The results include transverse momentum (Fig.~\ref{fig:ptmty}, \textit{top left}) and transverse mass ($\mt = \mkern-4mu\sqrt{\pt^2 + m_0^2}$) (Fig.~\ref{fig:ptmty}, \textit{top right}) spectra, rapidity spectra (Fig.~\ref{fig:ptmty}, \textit{bottom left}), and $K^{*}(892)^0$ mean multiplicities.

The transverse momentum and transverse mass spectra were obtained in the rapidity range $0 < \y < 1.5$. The solid lines in Fig.~\ref{fig:ptmty} (\textit{top left}, \textit{top right}) correspond to fitted exponential functions with inverse slope parameters: $T = 226 \pm 17~\mathrm{(stat)} \pm 19~\mathrm{(sys)}$ \MeV for 40\AGeVc, $T = 256 \pm 16~\mathrm{(stat)} \pm 13~\mathrm{(sys)}$ \MeV for 75\AGeVc, and $T = 284 \pm 21~\mathrm{(stat)} \pm 19~\mathrm{(sys)}$ \MeV for 150\AGeVc.
The fits were performed on the \pt spectra, and the same fit parameters were then used to draw solid lines on the \mt spectra. The deviations of the individual data points from the exponential fits remain below $2\sigma_\mathrm{tot}$ ($\sigma_\mathrm{tot}= \mkern-4mu \sqrt{\mathrm{stat}^2+\mathrm{sys}^2}$). The $T$ parameter increases with collision energy, which may be related to an increase in radial flow velocities.

The rapidity spectra, presented in Fig.~\ref{fig:ptmty} (\textit{bottom left}), were obtained in the transverse momentum range $0 < \pt < 1.5$ \GeVc, which at SPS energies covers almost the entire production of $K^{*}(892)^0$. The full symbols represent the measurements, and the open symbols were obtained by reflection around mid-rapidity. The shape of the $\dd n/\dd \y$ distribution at the top SPS beam momentum displays a significant difference with respect to those obtained at 40\A and 75\AGeVc.
Mean multiplicity was calculated as the sum of measured rapidity points (multiplied by bin widths) scaled under the assumption that the ratio between measured and unmeasured regions is the same in data and \EposLong (or \FTFPBERT when determining systematic uncertainty)~\cite{NA61SHINE:2021uyb, BKozlowski_PhD}. 
The mean multiplicities of $K^{*}(892)^0$ mesons in $0 < \pt < 1.5$ \GeVc are as follows: 
$\langle K^*(892)^0 \rangle = 1.41 \pm 0.13~\mathrm{(stat)} \pm 0.11~\mathrm{(sys)}$ for \ArSc at 40\AGeVc, 
$\langle K^*(892)^0 \rangle = 1.527 \pm 0.096~\mathrm{(stat)} \pm 0.18~\mathrm{(sys)}$ for \ArSc at 75\AGeVc, and 
$\langle K^*(892)^0 \rangle = 2.37 \pm 0.14~\mathrm{(stat)} \pm 0.16~\mathrm{(sys)}$ for \ArSc at 150\AGeVc.

\begin{figure}[h]
\centering
\includegraphics[width=0.47\textwidth]{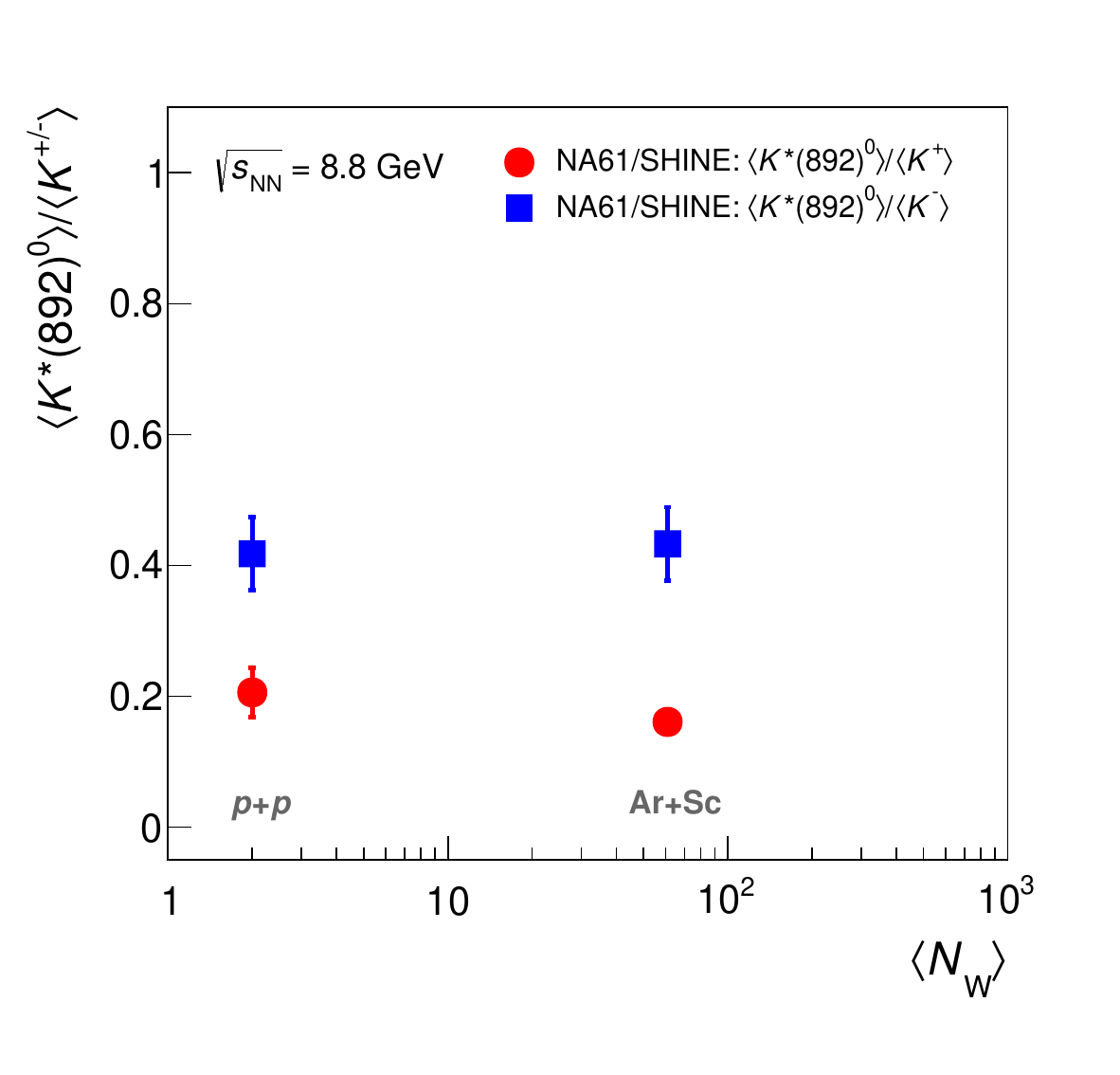}
\includegraphics[width=0.47\textwidth]{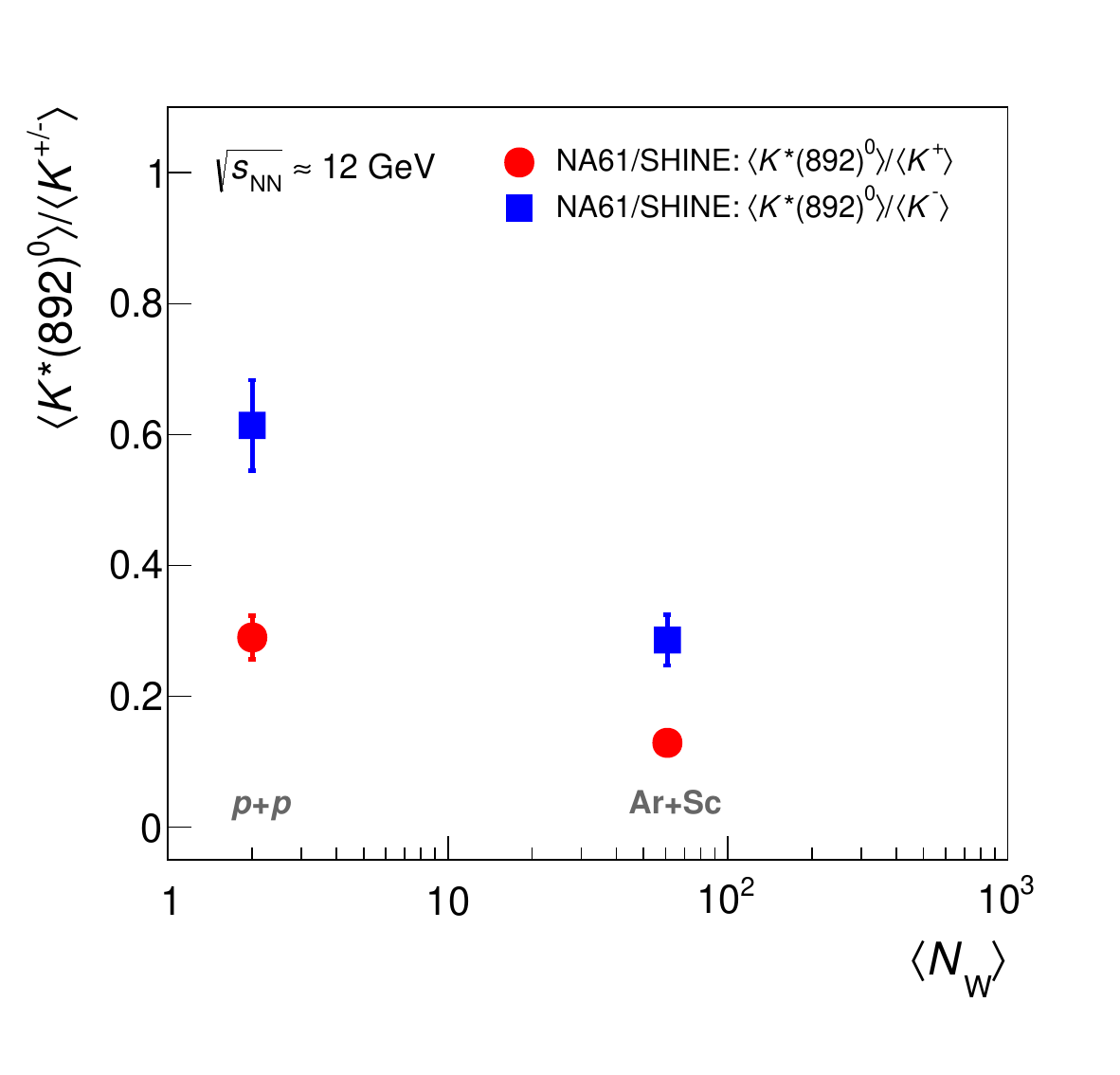}
\includegraphics[width=0.47\textwidth]{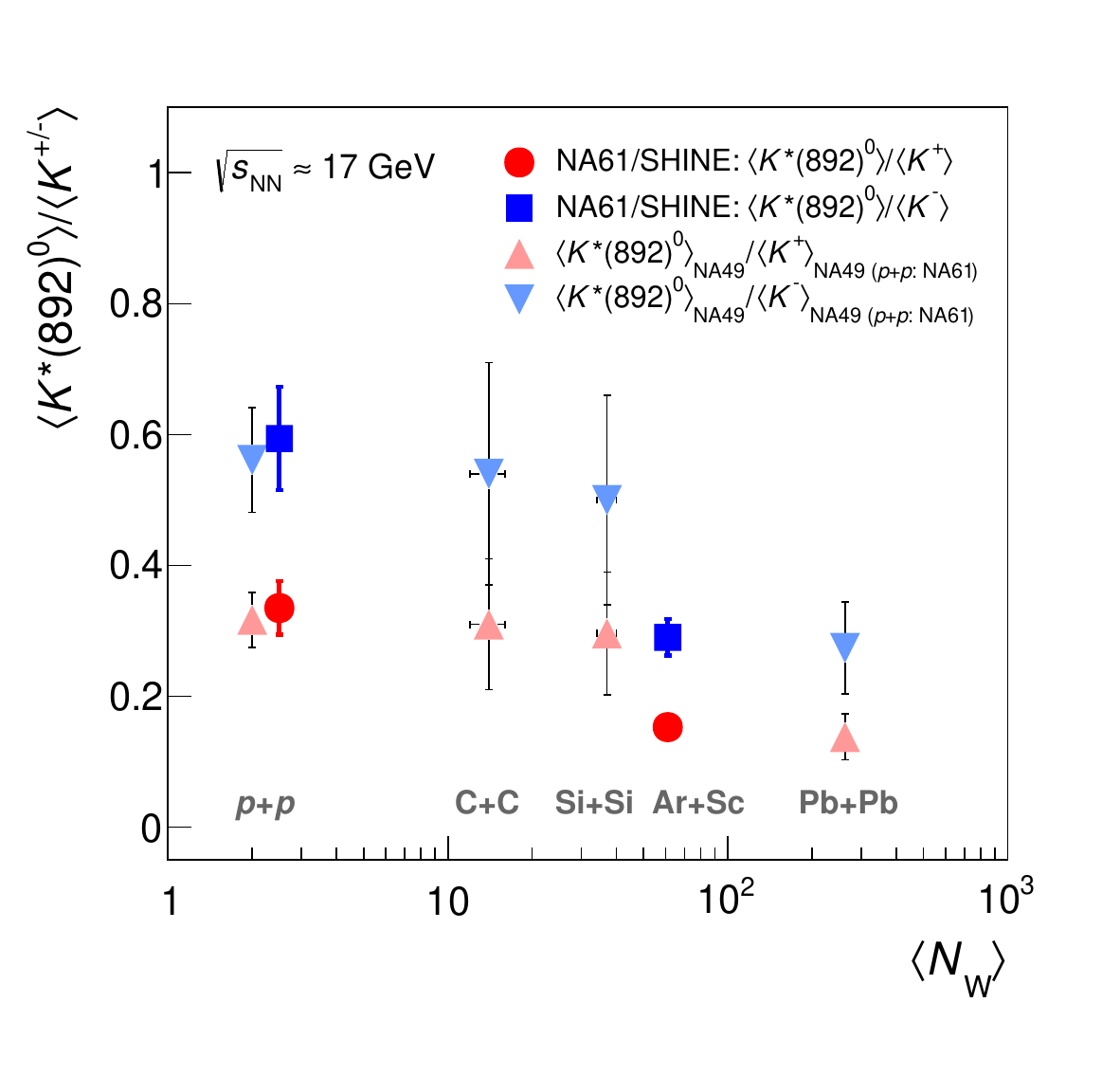}
\includegraphics[width=0.47\textwidth]{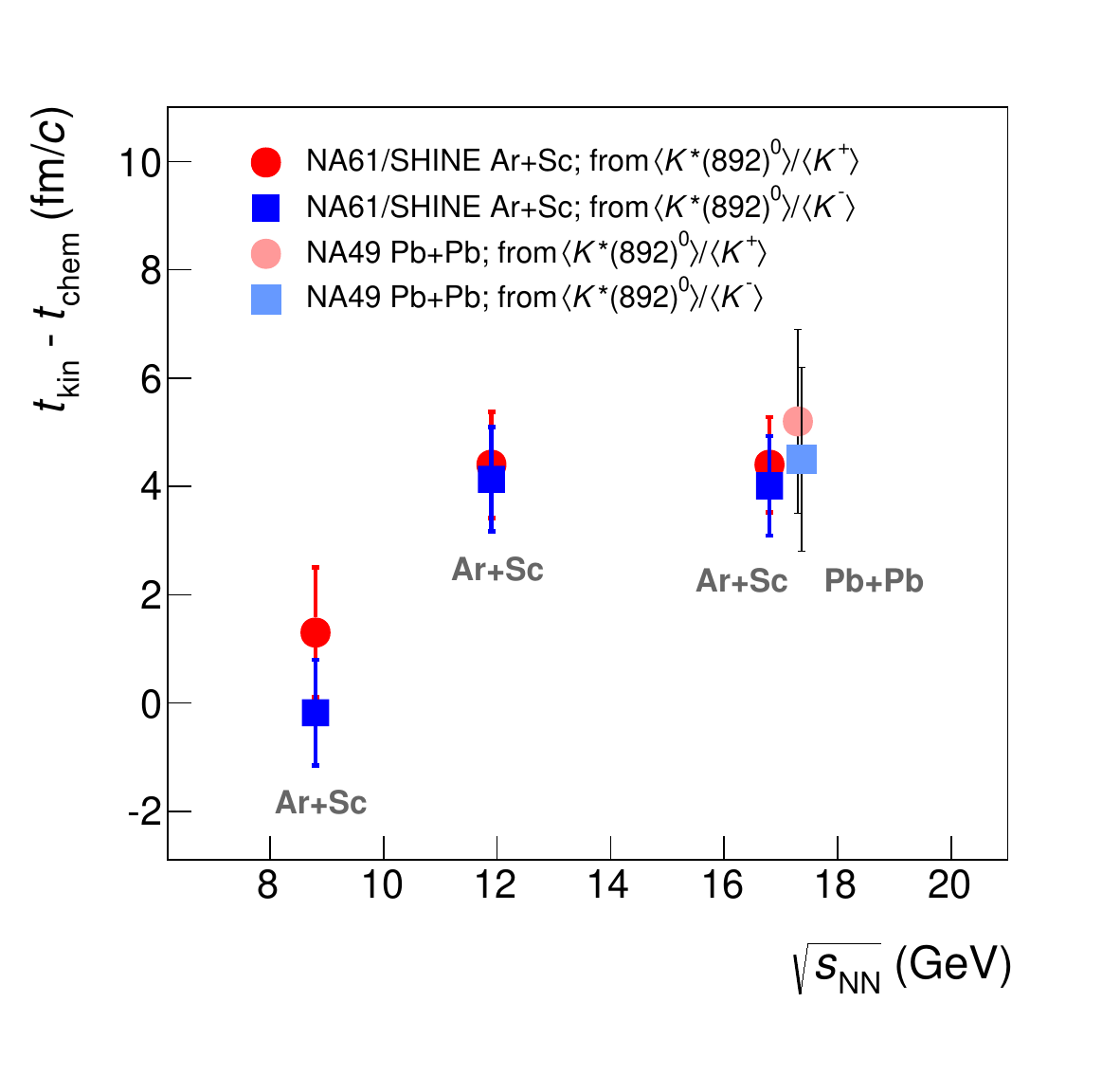}
\vspace{-0.4cm}
\caption{The system size dependencies ($\langle N_\mathrm{W} \rangle$ corresponds to the mean number of wounded nucleons) of $\langle K^{*}(892)^0 \rangle/\langle K^{+} \rangle $ and $\langle K^{*}(892)^0 \rangle/\langle K^{-} \rangle$ ratios ($\langle K^{+/-} \rangle$ on vertical axes denote $\langle K^{+} \rangle$ or $\langle K^{-} \rangle$) at \sNN $=$ 8.8~\GeV, $\sNN \approx$ 12 \GeV, and $\sNN \approx$ 17 \GeV (\textit{top} and \textit{bottom left}) as well as the energy dependence of time between freeze-outs (\textit{bottom right}). Plots include the \NASixtyOne results on $K^{*}(892)^0$ production in 0--10\% central \ArSc collisions as well as the published NA49 and \NASixtyOne results on $K^{*}(892)^0$~\cite{NA61SHINE:2021wba, NA61SHINE:2020czr, NA49:2011bfu} and $K^{+/-}$~\cite{NA61SHINE:2017fne, NA49:2004jzr, NA61SHINE:2023epu, NA49:2002pzu} yields. See Ref.~\cite{NA61SHINE:2020czr} for more details on compilation of $\langle K^{*}(892)^0 \rangle/\langle K^{+} \rangle $ and $\langle K^{*}(892)^0 \rangle/\langle K^{-} \rangle$ ratios at $\sNN \approx$ 17 \GeV. For all particle species ($K^{*}(892)^0$, $K^{+}$, $K^{-}$), the total uncertainties of mean multiplicities were calculated ($\mkern-4mu\sqrt{\mathrm{stat}^2+\mathrm{sys}^2}$) and then used to calculate the uncertainties of presented quantities.}
\label{fig:KstarKtime}
\end{figure}

The three panels of Fig.~\ref{fig:KstarKtime} (\textit{top} and \textit{bottom left}) show the system size dependencies of the $\langle K^{*}(892)^0 \rangle/\langle K^{+} \rangle $ and $\langle K^{*}(892)^0 \rangle/\langle K^{-} \rangle$ ratios at three SPS energies\footnote{Roughly the same value of $\langle K^{*}(892)^0 \rangle/\langle K^{-} \rangle$ in \pp and central \ArSc collisions at 8.8~\GeV and the surprisingly strong suppression of the $\langle K^{*}(892)^0 \rangle/\langle K^{-} \rangle$ and $\langle K^{*}(892)^0 \rangle/\langle K^{+} \rangle$ ratios from \pp to central \ArSc collisions at approximately 12 and 17~\GeV are not reproduced by the UrQMD hadronic transport model~\cite{Chabane:2025eir}. }. The decrease of the $K^{*}/K$ yield ratio from \pp to nucleus--nucleus collisions suggests that rescattering processes play an important role in a hadronic system produced in nucleus--nucleus collisions. Following Refs.~\cite{STAR:2004bgh, Markert:2002rw} the lifetime of such a hadronic system (time between freeze-outs) can be determined using the formula: 
\begin{equation}
\dfrac{K^*}{K}|_{\mathrm{kinetic}} = \dfrac{K^*}{K}|_{\mathrm{chemical}} \cdot e^{-\frac{\Delta t}{\tau}}\, ,  
\end{equation}
where $\dfrac{K^*}{K}|_{\mathrm{chemical}}$ is represented by the $K^*/K$ yield ratio in inelastic \pp collisions, $\dfrac{K^*}{K}|_{\mathrm{kinetic}}$ is represented by the $K^*/K$ yield ratio in nucleus--nucleus collisions, $\tau$ is the $K^*(892)^0$ lifetime taken as 4.17 fm/$c$~\cite{ParticleDataGroup:2024}, and $\Delta t$ is the time between chemical and kinetic freeze-outs in nucleus--nucleus collisions (in the $K^*$ rest frame). The obtained $\Delta t$ values, boosted by the Lorentz factor $\gamma = \mkern-4mu \sqrt{1+(\langle \pt \rangle / m_0 c)^2}$ (see Ref.~\cite{ALICE:2019xyr} for details), are presented in Fig.~\ref{fig:KstarKtime} (\textit{bottom right})\footnote{Mean transverse momenta of $K^*(892)^0$ mesons were obtained from fits to the \pt distributions~\cite{BKozlowski_PhD} and found to be
$\langle \pt \rangle = 685 \pm 34~\mathrm{(stat)} \pm 40~\mathrm{(sys)}$ \MeVc for 40\AGeVc, 
$\langle \pt \rangle = 745 \pm 31~\mathrm{(stat)} \pm 25~\mathrm{(sys)}$ \MeVc for 75\AGeVc, 
and $\langle \pt \rangle = 801 \pm 41~\mathrm{(stat)} \pm 38~\mathrm{(sys)}$ \MeVc for 150\AGeVc.
When calculating $\gamma$, their total uncertainties ($\mkern-4mu\sqrt{\mathrm{stat}^2+\mathrm{sys}^2}$) were used.}. 
The time between freeze-outs ($t_{\mathrm{kin}}-t_{\mathrm{chem}}=\gamma \Delta t$) is significantly larger than zero and similar for central \ArSc and \PbPb collisions at the highest energy ($\sNN \approx$ 17 \GeV). Moreover, it is also similar for \ArSc collisions at $\sNN \approx$ 12 \GeV and $\sNN \approx$ 17 \GeV. For \ArSc interactions at \sNN $=$ 8.8~\GeV, the observed time is smaller and, within uncertainties, remains consistent with zero. One should, however, stress that the possible effects of $K^{*}(892)^0$ regeneration before kinetic freeze-out are neglected in the above picture. Therefore, the estimated time intervals should rather be considered as lower limits of the time between freeze-outs.

The similarity of the times between freeze-outs in \ArSc collisions at $\sNN \approx 12~\GeV$ and $\sNN \approx 17~\GeV$ is consistent with RHIC Beam Energy Scan results, which exhibit comparable $K^*/K$ ratios and $\gamma \Delta t$ values over the energy range 7.7~\GeV $<$\sNN $\lesssim$ 20~\GeV in central Au+Au collisions~\cite{STAR:2022sir, STAR:2026kqj}. On the other hand, a rather monotonic decrease of the $K^*/K$ ratio is observed when moving from the most peripheral to the most central Au+Au collisions at these energies.

\vspace{-0.4cm}


\section{Summary}
\label{sec:summary}

The $K^{*}(892)^0$ resonance production has been measured in 0--10\% central \ArSc collisions at beam momenta 40\A, 75\A, and 150\AGeVc (\sNN $=$ 8.8, 11.9, 16.8 \GeV). Transverse momentum, transverse mass, and rapidity distributions together with mean multiplicities have been presented. The obtained $\langle K^{*}(892)^0 \rangle/\langle K^{+} \rangle$ and $\langle K^{*}(892)^0 \rangle /\langle K^{-} \rangle$ ratios at two higher energies show a decrease from \pp to \ArSc collisions, which is in agreement with rescattering effects present in nucleus--nucleus collisions. Such a decrease is not observed in \ArSc collisions at \sNN $=$ 8.8 \GeV. A rather surprising feature is the consistency of the time intervals between freeze-outs observed for the heavy \PbPb system and the intermediate-mass \ArSc system at the highest energy. The \NASixtyOne Collaboration plans to continue this analysis using \XeLa data. 
\clearpage

\section*{Acknowledgements}
We would like to thank the CERN EP, BE, HSE and EN Departments for the
strong support of NA61/SHINE.

This work was supported by
the Hungarian Scientific Research Fund (grant NKFIH 138136\slash137812\slash138152,
TKP2021-NKTA-64,
2025-1.1.5-NEMZ-KI-2025-00003),
the Polish Ministry of Science and Higher Education (WUT ID-UB), 
the National Science Centre Poland (grants
2018\slash 30\slash A\slash ST2\slash 00226, 
2018\slash 31\slash G\slash ST2\slash 03910, 
2020\slash 39\slash O\slash ST2\slash 00277, 
2021\slash 43\slash P\slash ST2\slash 03319, 
2023\slash 51\slash D\slash ST2\slash 02950), 
the Norwegian Financial Mechanism 2014--2021 (grant 2019\slash 34\slash H\slash ST2\slash 00585),
the Polish Minister of Education and Science (contract No. 2021\slash WK\slash 10),
the Polish Minister of Science and Higher Education (contract No. 2025\slash WK\slash 05),
the European Union's Horizon 2020 research and innovation programme under grant agreement No. 871072,
the Ministry of Education, Culture, Sports,
Science and Tech\-no\-lo\-gy, Japan, Grant-in-Aid for Sci\-en\-ti\-fic
Research (grants 18071005, 19034011, 19740162, 20740160 and 20039012,22H04943),
the German Research Foundation DFG (grants GA\,1480\slash8-1 and project 426579465),
the Bulgarian Ministry of Education and Science within the National
Roadmap for Research Infrastructures 2020--2027, contract No. D01-374/18.12.2020,
Swiss Nationalfonds Foundation (grant 200020\-117913/1),
ETH Research Grant TH-01\,07-3, National Science Foundation grant
PHY-2013228 and the Fermi National Accelerator Laboratory (Fermilab),
a U.S. Department of Energy, Office of Science, HEP User Facility
managed by Fermi Research Alliance, LLC (FRA), acting under Contract
No. DE-AC02-07CH11359 and the IN2P3-CNRS (France).\\


\bibliographystyle{include/na61Utphys}
\newlength{\bibitemsep}\setlength{\bibitemsep}{.0\baselineskip plus .00\baselineskip minus .00\baselineskip}
\newlength{\bibparskip}\setlength{\bibparskip}{0pt}
\let\oldthebibliography\thebibliography
\renewcommand\thebibliography[1]{%
  \oldthebibliography{#1}%
  \setlength{\parskip}{\bibitemsep}%
  \setlength{\itemsep}{\bibparskip}%
}
\bibliography{include/na61References}

\clearpage
{\Large The \NASixtyOne Collaboration}
\bigskip
\begin{sloppypar}

\noindent
{P.\;Adrich~\href{https://orcid.org/0000-0002-7019-5451}{\includegraphics[height=1.7ex]{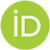}}\textsuperscript{\,13}},
{K.K.\;Allison~\href{https://orcid.org/0000-0002-3494-9383}{\includegraphics[height=1.7ex]{orcid-logo.png}}\textsuperscript{\,23}},
{M.\;Bajda~\href{https://orcid.org/0009-0005-8859-1099}{\includegraphics[height=1.7ex]{orcid-logo.png}}\textsuperscript{\,14}},
{Y.\;Balkova~\href{https://orcid.org/0000-0002-6957-573X}{\includegraphics[height=1.7ex]{orcid-logo.png}}\textsuperscript{\,13}},
{D.\;Battaglia~\href{https://orcid.org/0000-0002-5283-0992}{\includegraphics[height=1.7ex]{orcid-logo.png}}\textsuperscript{\,22}},
{M.\;Bielewicz~\href{https://orcid.org/0000-0001-8267-4874}{\includegraphics[height=1.7ex]{orcid-logo.png}}\textsuperscript{\,13}},
{A.\;Blondel~\href{https://orcid.org/0000-0002-1597-8859}{\includegraphics[height=1.7ex]{orcid-logo.png}}\textsuperscript{\,4}},
{M.\;Bogomilov~\href{https://orcid.org/0000-0001-7738-2041}{\includegraphics[height=1.7ex]{orcid-logo.png}}\textsuperscript{\,2}},
{Y.\;Bondar~\href{https://orcid.org/0000-0003-2773-9668}{\includegraphics[height=1.7ex]{orcid-logo.png}}\textsuperscript{\,11}},
{J.\;Brzychczyk~\href{https://orcid.org/0000-0001-5320-6748}{\includegraphics[height=1.7ex]{orcid-logo.png}}\textsuperscript{\,14}},
{M.\;Buryakov~\href{https://orcid.org/0009-0008-2394-4967}{\includegraphics[height=1.7ex]{orcid-logo.png}}\textsuperscript{\,20}},
{A.F.\;Camino\textsuperscript{\,25}},
{Y.D.\;Chandak~\href{https://orcid.org/0009-0009-2080-566X}{\includegraphics[height=1.7ex]{orcid-logo.png}}\textsuperscript{\,23}},
{M.\;Csan\'ad~\href{https://orcid.org/0000-0002-3154-6925}{\includegraphics[height=1.7ex]{orcid-logo.png}}\textsuperscript{\,7}},
{M.\;\'Cwiok~\href{https://orcid.org/0009-0007-1440-9113}{\includegraphics[height=1.7ex]{orcid-logo.png}}\textsuperscript{\,17}},
{T.\;Czopowicz~\href{https://orcid.org/0000-0003-1908-2977}{\includegraphics[height=1.7ex]{orcid-logo.png}}\textsuperscript{\,13}},
{C.\;Dalmazzone~\href{https://orcid.org/0000-0001-6945-5845}{\includegraphics[height=1.7ex]{orcid-logo.png}}\textsuperscript{\,4}},
{N.\;Davis~\href{https://orcid.org/0000-0003-3047-6854}{\includegraphics[height=1.7ex]{orcid-logo.png}}\textsuperscript{\,12}},
{A.\;Dmitriev~\href{https://orcid.org/0000-0001-7853-0173}{\includegraphics[height=1.7ex]{orcid-logo.png}}\textsuperscript{\,20}},
{P.~von\;Doetinchem~\href{https://orcid.org/0000-0002-7801-3376}{\includegraphics[height=1.7ex]{orcid-logo.png}}\textsuperscript{\,24}},
{W.\;Dominik~\href{https://orcid.org/0000-0001-7444-9239}{\includegraphics[height=1.7ex]{orcid-logo.png}}\textsuperscript{\,17}},
{J.\;Dumarchez~\href{https://orcid.org/0000-0002-9243-4425}{\includegraphics[height=1.7ex]{orcid-logo.png}}\textsuperscript{\,4}},
{R.\;Engel~\href{https://orcid.org/0000-0003-2924-8889}{\includegraphics[height=1.7ex]{orcid-logo.png}}\textsuperscript{\,5}},
{L.\;Fields~\href{https://orcid.org/0000-0001-8281-3686}{\includegraphics[height=1.7ex]{orcid-logo.png}}\textsuperscript{\,22}},
{Z.\;Fodor~\href{https://orcid.org/0000-0003-2519-5687}{\includegraphics[height=1.7ex]{orcid-logo.png}}\textsuperscript{\,6,18}},
{M.\;Friend~\href{https://orcid.org/0000-0003-4660-4670}{\includegraphics[height=1.7ex]{orcid-logo.png}}\textsuperscript{\,8}},
{M.\;Ga\'zdzicki~\href{https://orcid.org/0000-0002-6114-8223}{\includegraphics[height=1.7ex]{orcid-logo.png}}\textsuperscript{\,11}},
{K.E.\;Gollwitzer\textsuperscript{\,21}},
{V.\;Golovatyuk~\href{https://orcid.org/0009-0006-5201-0990}{\includegraphics[height=1.7ex]{orcid-logo.png}}\textsuperscript{\,20}},
{K.\;Grebieszkow~\href{https://orcid.org/0000-0002-6754-9554}{\includegraphics[height=1.7ex]{orcid-logo.png}}\textsuperscript{\,19}},
{P.G.\;Hurh~\href{https://orcid.org/0000-0002-9024-5399}{\includegraphics[height=1.7ex]{orcid-logo.png}}\textsuperscript{\,21}},
{S.\;Ilieva~\href{https://orcid.org/0000-0001-9204-2563}{\includegraphics[height=1.7ex]{orcid-logo.png}}\textsuperscript{\,2}},
{V.A.\;Kireyeu~\href{https://orcid.org/0000-0002-5630-9264}{\includegraphics[height=1.7ex]{orcid-logo.png}}\textsuperscript{\,20}},
{R.\;Kolesnikov~\href{https://orcid.org/0009-0006-4224-1058}{\includegraphics[height=1.7ex]{orcid-logo.png}}\textsuperscript{\,20}},
{D.\;Kolev~\href{https://orcid.org/0000-0002-9203-4739}{\includegraphics[height=1.7ex]{orcid-logo.png}}\textsuperscript{\,2}},
{Y.\;Koshio~\href{https://orcid.org/0000-0003-0437-8505}{\includegraphics[height=1.7ex]{orcid-logo.png}}\textsuperscript{\,9}},
{S.\;Kowalski~\href{https://orcid.org/0000-0001-9888-4008}{\includegraphics[height=1.7ex]{orcid-logo.png}}\textsuperscript{\,16}},
{T.\;Kowalski~\href{https://orcid.org/0000-0002-2550-1704}{\includegraphics[height=1.7ex]{orcid-logo.png}}\textsuperscript{\,13}},
{B.\;Koz{\l}owski~\href{https://orcid.org/0000-0001-8442-2320}{\includegraphics[height=1.7ex]{orcid-logo.png}}\textsuperscript{\,19}},
{A.\;Krasnoperov~\href{https://orcid.org/0000-0002-1425-2861}{\includegraphics[height=1.7ex]{orcid-logo.png}}\textsuperscript{\,20}},
{W.\;Kucewicz~\href{https://orcid.org/0000-0002-2073-711X}{\includegraphics[height=1.7ex]{orcid-logo.png}}\textsuperscript{\,15}},
{M.\;Kuchowicz~\href{https://orcid.org/0000-0003-3174-585X}{\includegraphics[height=1.7ex]{orcid-logo.png}}\textsuperscript{\,18}},
{P.\;Lasko~\href{https://orcid.org/0000-0003-1110-9522}{\includegraphics[height=1.7ex]{orcid-logo.png}}\textsuperscript{\,14}},
{A.\;L\'aszl\'o~\href{https://orcid.org/0000-0003-2712-6968}{\includegraphics[height=1.7ex]{orcid-logo.png}}\textsuperscript{\,6}},
{M.\;Lewicki~\href{https://orcid.org/0000-0002-8972-3066}{\includegraphics[height=1.7ex]{orcid-logo.png}}\textsuperscript{\,12}},
{G.\;Lykasov~\href{https://orcid.org/0000-0002-1544-6959}{\includegraphics[height=1.7ex]{orcid-logo.png}}\textsuperscript{\,20}},
{J.R.\;Lyon~\href{https://orcid.org/0009-0003-2579-8821}{\includegraphics[height=1.7ex]{orcid-logo.png}}\textsuperscript{\,24}},
{V.V.\;Lyubushkin~\href{https://orcid.org/0000-0003-0136-233X}{\includegraphics[height=1.7ex]{orcid-logo.png}}\textsuperscript{\,20}},
{M.\;Ma\'ckowiak-Paw{\l}owska~\href{https://orcid.org/0000-0003-3954-6329}{\includegraphics[height=1.7ex]{orcid-logo.png}}\textsuperscript{\,19}},
{B.\;Maksiak~\href{https://orcid.org/0000-0002-7950-2307}{\includegraphics[height=1.7ex]{orcid-logo.png}}\textsuperscript{\,13}},
{A.I.\;Malakhov~\href{https://orcid.org/0000-0001-8569-8409}{\includegraphics[height=1.7ex]{orcid-logo.png}}\textsuperscript{\,20}},
{A.\;Marcinek~\href{https://orcid.org/0000-0001-9922-743X}{\includegraphics[height=1.7ex]{orcid-logo.png}}\textsuperscript{\,12}},
{A.D.\;Marino~\href{https://orcid.org/0000-0002-1709-538X}{\includegraphics[height=1.7ex]{orcid-logo.png}}\textsuperscript{\,23}},
{T.\;Matulewicz~\href{https://orcid.org/0000-0003-2098-1216}{\includegraphics[height=1.7ex]{orcid-logo.png}}\textsuperscript{\,17}},
{V.\;Matveev~\href{https://orcid.org/0000-0002-2745-5908}{\includegraphics[height=1.7ex]{orcid-logo.png}}\textsuperscript{\,20}},
{G.L.\;Melkumov~\href{https://orcid.org/0009-0004-2074-6755}{\includegraphics[height=1.7ex]{orcid-logo.png}}\textsuperscript{\,20}},
{{\L}.\;Mik~\href{https://orcid.org/0000-0003-2712-6861}{\includegraphics[height=1.7ex]{orcid-logo.png}}\textsuperscript{\,15}},
{Y.\;Nagai~\href{https://orcid.org/0000-0002-1792-5005}{\includegraphics[height=1.7ex]{orcid-logo.png}}\textsuperscript{\,7,8}},
{R.\;Nagy~\href{https://orcid.org/0009-0004-4274-1832}{\includegraphics[height=1.7ex]{orcid-logo.png}}\textsuperscript{\,6}},
{T.\;Nakadaira~\href{https://orcid.org/0000-0003-4327-7598}{\includegraphics[height=1.7ex]{orcid-logo.png}}\textsuperscript{\,8}},
{S.\;Nishimori~\href{https://orcid.org/~0000-0002-1820-0938}{\includegraphics[height=1.7ex]{orcid-logo.png}}\textsuperscript{\,8}},
{A.\;Olivier~\href{https://orcid.org/0000-0003-4261-8303}{\includegraphics[height=1.7ex]{orcid-logo.png}}\textsuperscript{\,22}},
{V.\;Ozvenchuk~\href{https://orcid.org/0000-0002-7821-7109}{\includegraphics[height=1.7ex]{orcid-logo.png}}\textsuperscript{\,12}},
{O.\;Panova~\href{https://orcid.org/0000-0001-5039-7788}{\includegraphics[height=1.7ex]{orcid-logo.png}}\textsuperscript{\,12}},
{V.\;Paolone~\href{https://orcid.org/0000-0003-2162-0957}{\includegraphics[height=1.7ex]{orcid-logo.png}}\textsuperscript{\,25}},
{I.\;Pidhurskyi~\href{https://orcid.org/0000-0001-9916-9436}{\includegraphics[height=1.7ex]{orcid-logo.png}}\textsuperscript{\,11}},
{R.\;P{\l}aneta~\href{https://orcid.org/0000-0001-8007-8577}{\includegraphics[height=1.7ex]{orcid-logo.png}}\textsuperscript{\,14}},
{B.A.\;Popov~\href{https://orcid.org/0000-0001-5416-9301}{\includegraphics[height=1.7ex]{orcid-logo.png}}\textsuperscript{\,20,4}},
{B.\;P\'orfy~\href{https://orcid.org/0000-0001-5724-9737}{\includegraphics[height=1.7ex]{orcid-logo.png}}\textsuperscript{\,6,7}},
{D.\;Pszczel~\href{https://orcid.org/0000-0002-4697-6688}{\includegraphics[height=1.7ex]{orcid-logo.png}}\textsuperscript{\,13}},
{S.\;Pu{\l}awski~\href{https://orcid.org/0000-0003-1982-2787}{\includegraphics[height=1.7ex]{orcid-logo.png}}\textsuperscript{\,16}},
{L.\;Ren~\href{https://orcid.org/0000-0003-1709-7673}{\includegraphics[height=1.7ex]{orcid-logo.png}}\textsuperscript{\,23}},
{V.Z.\;Reyna~Ortiz~\href{https://orcid.org/0000-0002-7026-8198}{\includegraphics[height=1.7ex]{orcid-logo.png}}\textsuperscript{\,11}},
{D.\;R\"ohrich\textsuperscript{\,10}},
{M.\;Roth~\href{https://orcid.org/0000-0003-1281-4477}{\includegraphics[height=1.7ex]{orcid-logo.png}}\textsuperscript{\,5}},
{{\L}.\;Rozp{\l}ochowski~\href{https://orcid.org/0000-0003-3680-6738}{\includegraphics[height=1.7ex]{orcid-logo.png}}\textsuperscript{\,12}},
{M.\;Rumyantsev~\href{https://orcid.org/0000-0001-8233-2030}{\includegraphics[height=1.7ex]{orcid-logo.png}}\textsuperscript{\,20}},
{A.\;Rustamov~\href{https://orcid.org/0000-0001-8678-6400}{\includegraphics[height=1.7ex]{orcid-logo.png}}\textsuperscript{\,1}},
{M.\;Rybczy\'nski~\href{https://orcid.org/0000-0002-3638-3766}{\includegraphics[height=1.7ex]{orcid-logo.png}}\textsuperscript{\,11}},
{A.\;Rybicki~\href{https://orcid.org/0000-0003-3076-0505}{\includegraphics[height=1.7ex]{orcid-logo.png}}\textsuperscript{\,12}},
{D.\;Rybka~\href{https://orcid.org/0000-0002-9924-6398}{\includegraphics[height=1.7ex]{orcid-logo.png}}\textsuperscript{\,13}},
{K.\;Sakashita~\href{https://orcid.org/0000-0003-2602-7837}{\includegraphics[height=1.7ex]{orcid-logo.png}}\textsuperscript{\,8}},
{K.\;Schmidt~\href{https://orcid.org/0000-0002-0903-5790}{\includegraphics[height=1.7ex]{orcid-logo.png}}\textsuperscript{\,16}},
{P.\;Semeniuk~\href{https://orcid.org/0009-0006-8755-711X}{\includegraphics[height=1.7ex]{orcid-logo.png}}\textsuperscript{\,15}},
{H.\;Shah~\href{https://orcid.org/0000-0002-3021-6641}{\includegraphics[height=1.7ex]{orcid-logo.png}}\textsuperscript{\,12}},
{U.A.\;Shah~\href{https://orcid.org/0000-0002-9315-1304}{\includegraphics[height=1.7ex]{orcid-logo.png}}\textsuperscript{\,11}},
{Y.\;Shiraishi~\href{https://orcid.org/0000-0002-0132-3923}{\includegraphics[height=1.7ex]{orcid-logo.png}}\textsuperscript{\,9}},
{A.\;Shukla~\href{https://orcid.org/0000-0003-3839-7229}{\includegraphics[height=1.7ex]{orcid-logo.png}}\textsuperscript{\,24}},
{M.\;S{\l}odkowski~\href{https://orcid.org/0000-0003-0463-2753}{\includegraphics[height=1.7ex]{orcid-logo.png}}\textsuperscript{\,19}},
{P.\;Staszel~\href{https://orcid.org/0000-0003-4002-1626}{\includegraphics[height=1.7ex]{orcid-logo.png}}\textsuperscript{\,14}},
{G.\;Stefanek~\href{https://orcid.org/0000-0001-6656-9177}{\includegraphics[height=1.7ex]{orcid-logo.png}}\textsuperscript{\,11}},
{J.\;Stepaniak~\href{https://orcid.org/0000-0003-2064-9870}{\includegraphics[height=1.7ex]{orcid-logo.png}}\textsuperscript{\,13}},
{T.\;\v{S}u\v{s}a~\href{https://orcid.org/0000-0001-7430-2552}{\includegraphics[height=1.7ex]{orcid-logo.png}}\textsuperscript{\,3}},
{{\L}.\;\'Swiderski~\href{https://orcid.org/0000-0001-5857-2085}{\includegraphics[height=1.7ex]{orcid-logo.png}}\textsuperscript{\,13}},
{J.\;Szewi\'nski~\href{https://orcid.org/0000-0003-2981-9303}{\includegraphics[height=1.7ex]{orcid-logo.png}}\textsuperscript{\,13}},
{R.\;Szukiewicz~\href{https://orcid.org/0000-0002-1291-4040}{\includegraphics[height=1.7ex]{orcid-logo.png}}\textsuperscript{\,18}},
{A.\;Tefelska~\href{https://orcid.org/0000-0002-6069-4273}{\includegraphics[height=1.7ex]{orcid-logo.png}}\textsuperscript{\,19}},
{D.\;Tefelski~\href{https://orcid.org/0000-0003-0802-2290}{\includegraphics[height=1.7ex]{orcid-logo.png}}\textsuperscript{\,19}},
{V.\;Tereshchenko~\href{https://orcid.org/0000-0001-8996-2254}{\includegraphics[height=1.7ex]{orcid-logo.png}}\textsuperscript{\,20}},
{R.\;Tsenov~\href{https://orcid.org/0000-0002-1330-8640}{\includegraphics[height=1.7ex]{orcid-logo.png}}\textsuperscript{\,2}},
{L.\;Turko~\href{https://orcid.org/0000-0002-5474-8650}{\includegraphics[height=1.7ex]{orcid-logo.png}}\textsuperscript{\,18}},
{M.\;Unger~\href{https://orcid.org/0000-0002-7651-0272}{\includegraphics[height=1.7ex]{orcid-logo.png}}\textsuperscript{\,5}},
{M.\;Urbaniak~\href{https://orcid.org/0000-0002-9768-030X}{\includegraphics[height=1.7ex]{orcid-logo.png}}\textsuperscript{\,16}},
{D.\;Veberi\v{c}~\href{https://orcid.org/0000-0003-2683-1526}{\includegraphics[height=1.7ex]{orcid-logo.png}}\textsuperscript{\,5}},
{O.\;Vitiuk~\href{https://orcid.org/0000-0002-9744-3937}{\includegraphics[height=1.7ex]{orcid-logo.png}}\textsuperscript{\,18}},
{A.\;Wickremasinghe~\href{https://orcid.org/0000-0002-5325-0455}{\includegraphics[height=1.7ex]{orcid-logo.png}}\textsuperscript{\,21}},
{K.\;Witek~\href{https://orcid.org/0009-0004-6699-1895}{\includegraphics[height=1.7ex]{orcid-logo.png}}\textsuperscript{\,15}},
{K.\;W\'ojcik~\href{https://orcid.org/0000-0002-8315-9281}{\includegraphics[height=1.7ex]{orcid-logo.png}}\textsuperscript{\,16}},
{A.\;Zaitsev~\href{https://orcid.org/0000-0003-4711-9925}{\includegraphics[height=1.7ex]{orcid-logo.png}}\textsuperscript{\,20}},
{E.\;Zherebtsova~\href{https://orcid.org/0000-0002-1364-0969}{\includegraphics[height=1.7ex]{orcid-logo.png}}\textsuperscript{\,18}}, and
{E.D.\;Zimmerman~\href{https://orcid.org/0000-0002-6394-6659}{\includegraphics[height=1.7ex]{orcid-logo.png}}\textsuperscript{\,23}}

\end{sloppypar}

\noindent
\textsuperscript{1}~National Nuclear Research Center, Baku, Azerbaijan\\
\textsuperscript{2}~Faculty of Physics, University of Sofia, Sofia, Bulgaria\\
\textsuperscript{3}~Ru{\dj}er Bo\v{s}kovi\'c Institute, Zagreb, Croatia\\
\textsuperscript{4}~LPNHE, Sorbonne University, CNRS/IN2P3, Paris, France\\
\textsuperscript{5}~Karlsruhe Institute of Technology, Karlsruhe, Germany\\
\textsuperscript{6}~HUN-REN Wigner Research Centre for Physics, Budapest, Hungary\\
\textsuperscript{7}~E\"otv\"os Lor\'and University, Budapest, Hungary\\
\textsuperscript{8}~Institute for Particle and Nuclear Studies, Tsukuba, Japan\\
\textsuperscript{9}~Okayama University, Okayama, Japan\\
\textsuperscript{10}~University of Bergen, Bergen, Norway\\
\textsuperscript{11}~Jan Kochanowski University, Kielce, Poland\\
\textsuperscript{12}~Institute of Nuclear Physics, Polish Academy of Sciences, Cracow, Poland\\
\textsuperscript{13}~National Centre for Nuclear Research, Warsaw, Poland\\
\textsuperscript{14}~Jagiellonian University, Cracow, Poland\\
\textsuperscript{15}~AGH University of Krakow, Cracow, Poland\\
\textsuperscript{16}~University of Silesia, Katowice, Poland\\
\textsuperscript{17}~University of Warsaw, Warsaw, Poland\\
\textsuperscript{18}~University of Wroc{\l}aw,  Wroc{\l}aw, Poland\\
\textsuperscript{19}~Warsaw University of Technology, Warsaw, Poland\\
\textsuperscript{20}~Joint Institute for Nuclear Research, Dubna, International Organization\\
\textsuperscript{21}~Fermilab, Batavia, USA\\
\textsuperscript{22}~University of Notre Dame, Notre Dame, USA\\
\textsuperscript{23}~University of Colorado, Boulder, USA\\
\textsuperscript{24}~University of Hawaii at Manoa, Honolulu, USA\\
\textsuperscript{25}~University of Pittsburgh, Pittsburgh, USA\\

\end{document}